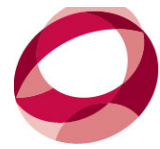

# Research Agenda in Intelligent Infrastructure to Enhance Disaster Management, Community Resilience and Public Safety

Michael Dunaway
University of Louisiana at Lafayette

Robin Murphy
Texas A&M University

Nalini Venkatasubramanian
University of California Irvine

Leysia Palen
University of Colorado at Boulder

Daniel Lopresti
Lehigh University

*Challenge and Opportunity*

Modern societies can be understood as the intersection of four interdependent systems: (1) the natural environment of geography, climate and weather; (2) the built environment of cities, engineered systems, and physical infrastructure; (3) the social environment of human populations, communities and socio-economic activities; and (4) an information ecosystem that overlays the other three domains and provides the means for understanding, interacting with, and managing the relationships between the natural, built, and human environments.

As the nation and its communities become more connected, networked and technologically sophisticated, new challenges and opportunities arise that demand a rethinking of current approaches to public safety and emergency management.[1] Addressing the current and future challenges requires an equally sophisticated program of research, technology development, and strategic planning. The design and integration of intelligent infrastructure—including embedded sensors, the Internet of Things (IoT), advanced wireless information technologies, real-time data capture and analysis, and machine-learning-based decision support—holds the potential to greatly enhance public safety, emergency management, disaster recovery, and overall community resilience, while addressing new and emerging threats to public safety and security. Ultimately, the objective of this program of research and development is to *save lives, reduce risk and disaster impacts, permit efficient use of material and social resources, and protect quality of life and economic stability* across entire regions.

Recent advances in emergency management and public safety have been characterized by the integration of technology applications with data and information systems, accompanied by organizational and cultural change. Two notable examples include the FEMA "whole of community" approach to emergency management that combines a formal Incident Command System with data visualization and decision systems, and establishes a National Disaster Recovery Framework targeting community recovery—and not simply emergency response—as the objective of disaster preparedness; and the Department of Justice "Community Policing" initiative that combines technology enabled police forces, integrated security systems and data analytics with education, engagement, and empowerment of the civil population. Similar initiatives in collaborative research and development of intelligent infrastructure and technologies are represented by programs like the NIST and National Science Foundation Smart & Connected Communities programs that emphasize joint efforts involving research universities and laboratories, commercial entities and technology developers, end-users such as first responder groups and local government agencies, aligned with federal funding agencies and non-profits.

At the leading edge of these initiatives is the collection, integration, management and analysis of an increasingly complex web of multi-modal data and digital information originating from mobile, fixed, and embedded sources—the information infrastructure that overlays and connects the natural, built and social environments. The now commonplace access to and use of cutting-edge information technologies by private citizens, commercial enterprises, and government agencies has made possible the development of a networked intelligent infrastructure with the capacity to collect multi-source data, and conduct near real-time analyses that integrate social, economic and physical dimensions. These capabilities hold the promise to enhance critical decision-making in public safety and emergency management through shared situational awareness, predictive analytics, and organizational learning across agencies.

However, realizing this potential will require continuing investment in research and development—targeting both physical and social infrastructures—to define technological requirements, analytic methodologies, social policies and strategies to guide development and integration of intelligent infrastructure systems. The success of this endeavor will depend on continuing partnerships among university research centers, federal laboratories, private sector research and development, and support of federal funding agencies.

---

[1] Mynatt et al. (2017) "A National Research Agenda for Intelligent Infrastructure" CCC Led Whitepapers http://cra.org/ccc/resources/ccc-led-whitepapers/, last accessed April 12, 2017.

*Benefits*

The nation currently faces both a strategic challenge and an opportunity brought about by the intersection of three major trends: (1) increasing technological sophistication of urban centers and the connectedness of both physical and social systems; (2) dramatic changes in hazards and risks at national, regional and local levels, to include the enduring threat of mass-casualty terrorism, and the increasing frequency of severe weather events and impact on communities and populations; and (3) the need for substantial revitalization of the nation's critical infrastructure systems, to include enhanced abilities to detect, prevent and mitigate cyber-security threats to critical high-reliability systems. These three trends highlight an urgent need for effects-based research into intelligent infrastructure systems that can achieve long-term benefits in resilience, sustainability, and disaster resistance, while providing the information backbone for enhanced data collection, analysis, and measurement of effects and outcomes.

Research in intelligent infrastructure will save lives, improve disaster resilience and enhance quality of life, accelerate social and economic recovery and care for victims in disaster-affected regions, and ensure continuity of governance. It will also create new jobs in both public and private sectors for individuals who have a high comfort level with computers, robotics, and networks—such as veterans and displaced knowledge workers. Moreover, research investment will preserve America's existing high-tech jobs, and maintain our edge in technology innovation, rather than allowing other countries (or asymmetric agents) to capitalize on, re-sell, or co-opt US industrial creativity. Finally, a national program for investment in intelligent infrastructure can achieve dramatic economies of scale and reduce long-term national debt. For example, the National Flood Insurance Program has spent $51B in claims from 1979-2014 and is currently $24B in debt. The nation spends billions of dollars annually to suppress catastrophic wildfires, which consume millions of acres each year. A single major hurricane or tornado can claim hundreds of lives and cause billions in damage. Intelligent infrastructure technologies such as computer models taught through machine learning and calibrated on big data, combined with large-scale instrumentation including ground-based sensors, streaming video from unmanned aerial vehicles, and satellite imagery, could significantly reduce the social and economic costs of such disasters.

*Research Agenda for Public Safety and Resilience*

Dramatic advances in disaster computation are possible with a long-term research investment to facilitate and sustain an organic research community. A roadmap would create a common ground for stakeholders and researchers and serve as the basis for consensus on high-priority, high-payoff computing capabilities, lower the bar of entry for the next generation of researchers, and encourage adoption of technologies by government agencies, operational services, and non-government organizations. A broad portfolio is envisioned comprised of a mix of traditional small, medium, and large grants managed through a network of regional university-based centers that would connect local stakeholders with researchers and provide testbeds of sufficiently high fidelity and scale to support formative experimentation and evaluation of progress. Priority should be placed on the following areas in intelligent infrastructure:

*Sensing, and Data Collection*
- Sensing research (particularly computer vision) to advance capabilities in environmental monitoring and interpretation of imagery from mobile perceptive devices (e.g., unmanned aerial, ground, or marine systems, smart phones, etc.) and associated video feeds from social media networks.
- Methods to enable resilient integration, operation, and security of next generation IoT devices and technologies, combining in-situ and mobile sensing deployed in the field.
- Coordination of air and ground assets for information gathering, and command and control of deployed responders and organizations engaged in active intervention in disaster scenarios.

*Communication and Coordination*
- Seamless multi-network provisioning utilizing any and all existing wired and wireless infrastructures, new long-range wireless technologies, 5G, and public safety networks cooperatively managed for reliable exchange of crisis-related information.
- Design of sense-analyze-respond systems to enhance first responder and agency situational awareness and support near real-time decision making.

- o Impact of system degradation and failures within interdependent infrastructures to develop analytics for response processes within closely coupled systems and avert cascading casualties.
- o Integration and coordination of multi-team systems, i.e., multi-disciplinary teams of first responder groups including firefighters, law enforcement, emergency medical, search and rescue, and emergency management that must share a common operating picture and coordinate operations across geography in dynamic, high-stress environments.
- o Research into Software Defined Networks for emergency management communications to establish and manage priority among disaster-related data transmitted via both standard wireless carriers and gigabit-capacity networks that may become degraded or overloaded (particularly with streaming video) during emergency conditions.

*Big Data Modeling frameworks, Analytics and Tools for Disaster Prediction and Management*
- o Probabilistic modeling of complex events to develop predictive analytics and enhance the capabilities for appropriate and adaptive response, and to refine response planning.
- o Multilevel, multiscale modeling methods for understanding factors that contribute to or undermine community resilience, and permits the capture and visualization of data elements reflecting different aspects of a community, from physical geography to built infrastructure to activities, entities, events, and processes on the infrastructure, and the impact on overall community resilience metrics. Such multiscale modeling techniques would be designed to capture entities and events in the framework at different levels of abstraction.
- o Research into protocols and methods for ensuring both reliability and privacy of data collection and analytics during emergency situations, disasters, and crises.

*Social Computing, Human Factors, and the Information Infrastructure*
- o Advanced information infrastructure that integrates information technologies with new analytic approaches to yield new data for new situations, applications and communities. This aim reflects a multidisciplinary approach that combines computing and social science with knowledge of disaster characteristics to enable the capture of information generated by the affected populations for conditions and crises of previously unknown character (e.g., widespread pandemic, nuclear terrorist incident, or a regional-scale earthquake).
- o Research into disaster resilient communications technologies and data management strategies to manage uncertainty and respond creatively across disaster warning, response, and recovery. This effort includes forecast data and sensor data, as well as data from online participation in social media and peer production platforms.
- o Research into the integration of social network analysis with geographic information systems (GIS) mapping to permit visual representation of human populations (the community civic infrastructure) as is currently done for mapping community critical infrastructures.
- o Research into learning systems and predictive analytics to aid decision-making during large-scale, mass-casualty events that affect populations and communities differentially, with situations changing quickly over time. In these dynamic situations, the data produced through social media by people affected by disasters are hyper-localized and hyper-temporalized, as are the answers and assistance sought by affected populations.
- o Development of algorithms for anticipatory information management based on social media queries and data requests originating from disaster affected populations;
- o Methods for reliably incorporating human input via social media and other sources to augment situational awareness with citizen perspectives of ground truth during incidents.
- o Natural interfaces and tools that enhance human capabilities to ask questions in a high-level manner that facilitate making decisions during extreme events.

***Summary: Intelligent Infrastructure for Research in Disaster Management, Resilience, and Public Safety***

The preceding highlights a significant research agenda for development, deployment, and integration of Intelligent Infrastructure to enhance community resilience, public safety, and disaster management in the emerging environment of smart and connected communities. An equal imperative exists for building and sustaining an *Intelligent Research Infrastructure* to advance capabilities in data acquisition and dissemination, analytics and decision-systems, human factors and social science, and to ensure continuing development of the next generation of researchers and technology developers. Continuing investment in university research centers, private sector enterprise, and government laboratories and agencies is a critical component of this national intelligent infrastructure.

*This material is based upon work supported by the National Science Foundation under Grant No. (1136993). Any opinions, findings, and conclusions or recommendations expressed in this material are those of the authors and do not necessarily reflect the views of the National Science Foundation.*